%% file: SMACS2031MUSE_MODEL.tex
\documentclass[useAMS,usenatbib]{mn2e}

\usepackage{graphicx}
\usepackage{txfonts}
\usepackage{amssymb,alltt}
\usepackage{color}
\usepackage{hyperref}
\usepackage{supertabular}

\hypersetup{colorlinks, linkcolor=blue}

\title[MUSE Strong-Lensing Analysis of SMACSJ2031]{MUSE observations of the lensing cluster SMACSJ2031.8-4036: 
new constraints on the mass distribution in the cluster core}
\author[Richard et al. 2014]
{Johan Richard$^{1}$\thanks{E-mail:
johan.richard@univ-lyon1.fr}, Vera Patricio$^{1}$, Johany Martinez$^{1}$, 
Roland Bacon$^{1}$, Benjamin Cl\'ement$^{1}$, 
\newauthor
Peter Weilbacher$^{2}$, 
Kurt Soto$^{3}$, Lutz Wisotzki$^{2}$, Jo\"el Vernet$^{4}$, Roser Pello$^{5}$, Joop Schaye$^{6}$, 
\newauthor
Monica Turner $^{6}$, Thomas Martinsson$^{6}$\\
\\
$^{1}$CRAL, Observatoire de Lyon, Universit\'e Lyon 1, 9 Avenue Ch. Andr\'e, 69561 Saint Genis Laval Cedex, France\\
$^{2}$ AIP, Leibniz-Institut f\"ur Astrophysik Potsdam (AIP) An der Sternwarte 16, D-14482 Potsdam\\
$^{3}$ ETH Zurich, Institute of Astronomy, HIT J 12.3 , Wolfgang-Pauli-Str. 27, CH-8093 Zurich, Switzerland\\
$^{4}$ European Southern Observatory, Karl-Schwarzschild Str. 2,  85748 Garching bei Muenchen, Germany\\
$^{5}$ Institut de Recherche en Astrophysique et Planetologie, Observatoire Midi-Pyr\'en\'ees, 14, Avenue Edouard Belin, F-31400 Toulouse, France\\
$^{6}$ Leiden Observatory, P.O. Box 9513, NL-2300 RA Leiden, Netherlands\\
}
\begin{document}

\date{Accepted 2014 September 08. Received 2014 September 07; in original form 2014 August 15.}

\pagerange{\pageref{firstpage}--\pageref{lastpage}} \pubyear{2014}

\maketitle

\label{firstpage}

\begin{abstract}
We present new observations of the lensing cluster SMACSJ2031.8-4036 obtained with the MUSE integral field spectrograph 
as part of its commissioning on the Very Large Telescope. By providing medium-resolution spectroscopy over the full 4750-9350 \AA\ 
domain and a 1x1 arcmin$^2$ field of view, MUSE is ideally suited for identifying lensed galaxies in the cluster core, in particular multiple-imaged systems. We perform a redshift analysis of all sources in the datacube and identify a total of 12 systems ranging 
from $z=1.46$ to $z=6.4$, with all images of each system confirmed by a spectroscopic redshift. This allows us to accurately constrain the 
cluster mass profile in this region. We foresee that future MUSE observations of cluster cores should help us discover very faint 
Lyman-$\alpha$ emitters thanks to the strong magnification and the high sensitivity of this instrument.
\end{abstract}



\section{Introduction}

Galaxy cluster cores contain large concentrations of dark matter, and in such act as 
powerful gravitational lenses. At its strongest, the deflection of light rays 
from distant galaxies can create multiple images of a single source \citep{2011A&ARv..19...47K}, and the location 
of such multiples can be used to infer the total mass distribution in cluster cores \citep{2010MNRAS.404..325R,2012ApJ...749...97Z}.

Since the discovery of the first lensed systems in cluster cores, dedicated 
whole-sky cluster searches such as the MAssive Cluster Survey (MACS, \citealt{2001ApJ...553..668E,2010MNRAS.407...83E}), followed 
up by multi-band Hubble imaging, have provided a large number of multiple systems in individual 
lensing clusters \citep{Zitrin2011,2012A&A...544A..71L}. The largest number of multiple systems so far have been found in 
the clusters Abell 1689 (34 systems, \citealt{2007ApJ...668..643L}), and the recent Frontier fields 
clusters (\citealt{Richard2014,Jauzac2014a,Jauzac2014b} with 68 systems for MACSJ0416.1-2403).

An important part of the mass modelling relies on getting accurate spectroscopic 
redshifts for the lensed sources, to completely confirm the lensing system and to avoid degeneracies  
in the mass normalisation. As the typical magnitude range for lensed sources found with Hubble 
is $\sim$23-25 AB \citep{Richard09,Jauzac2014a}, deep spectroscopy is required. Large spectroscopic 
campaigns have been performed with multi-object spectrographs (\citealt{2014arXiv1405.0222J,
2014arXiv1407.7866G}), but they rely on a preselection based on deep multi-band Hubble images 
and are quite inefficient since all multiple images are crowded in the cluster core.

The advent of the Multi Unit Spectroscopic Explorer (MUSE, \citealt{muse}), recently commissioned 
on the Very Large Telescope, may well set a new benchmark in the study of 
galaxy cluster cores. By providing integral field optical spectroscopy over a sky area of 1$\times$1 
arcmin$^2$, MUSE perfectly catches the full locus of multiple images in a moderately massive cluster. 
MUSE acts as a powerful redshift machine to identify faint emission lines of high equivalent width not 
appearing in classical continuum imaging, in particular Lyman-$\alpha$ emitters at $2.8<z<6.7$ probing 
the high-redshift Universe. 

Here, we present the first results obtained on deep pointed MUSE observations 
towards a massive cluster, SMACSJ2031.8-4036. We perform a redshift measurement of 
all sources in the MUSE field of view and identify multiply-imaged systems based on  
spectroscopic redshifts, the majority of which are not detected with short exposure Hubble images. 
This allows us to improve the cluster mass model in this region very significantly. Throughout this 
\textit{Letter}, we adopt a $\Lambda$-CDM cosmology with $\Omega_\Lambda$=0.7, 
$\Omega_{\rm M}$=0.3 and $H=70$ km/s/Mpc. At the redshift of the cluster $z=0.331$, 
1$\arcsec$ on sky corresponds to a physical distance of 4.763 kpc.

\section{Observations}
\label{observations}

SMACSJ2031.8-4036 ($\alpha$=20:31:47.843, $\delta$=-40:36:54.76, $z=0.331$, \citealt{Christensen2012}, hereafter SMACSJ2031) 
has been observed with the Hubble Space Telescope as part of snapshot programs 
12166 and 12884 (PI: Ebeling), using  the ACS and WFC3 cameras. The ACS data were taken 
in the F606W and F814W filters for 1.2 and 1.44 ksec respectively, and the WFC3 data were 
taken in F110W and F140W for 0.7 ksec each. The data were reduced using 
multidrizzle \citep{Koekemoer} and the astrometry tied to the USNO catalog.

The Hubble images reveal a bimodal cluster, with two distinct concentrations of 
cluster galaxies located 1.2 arcmin apart. The HST images are centred on the NW subcluster, 
and therefore the WFC3 data only partially cover the SE subcluster ($\sim$40\% of 
the MUSE field of view described below).

\begin{figure}
\includegraphics[width=0.5\textwidth]{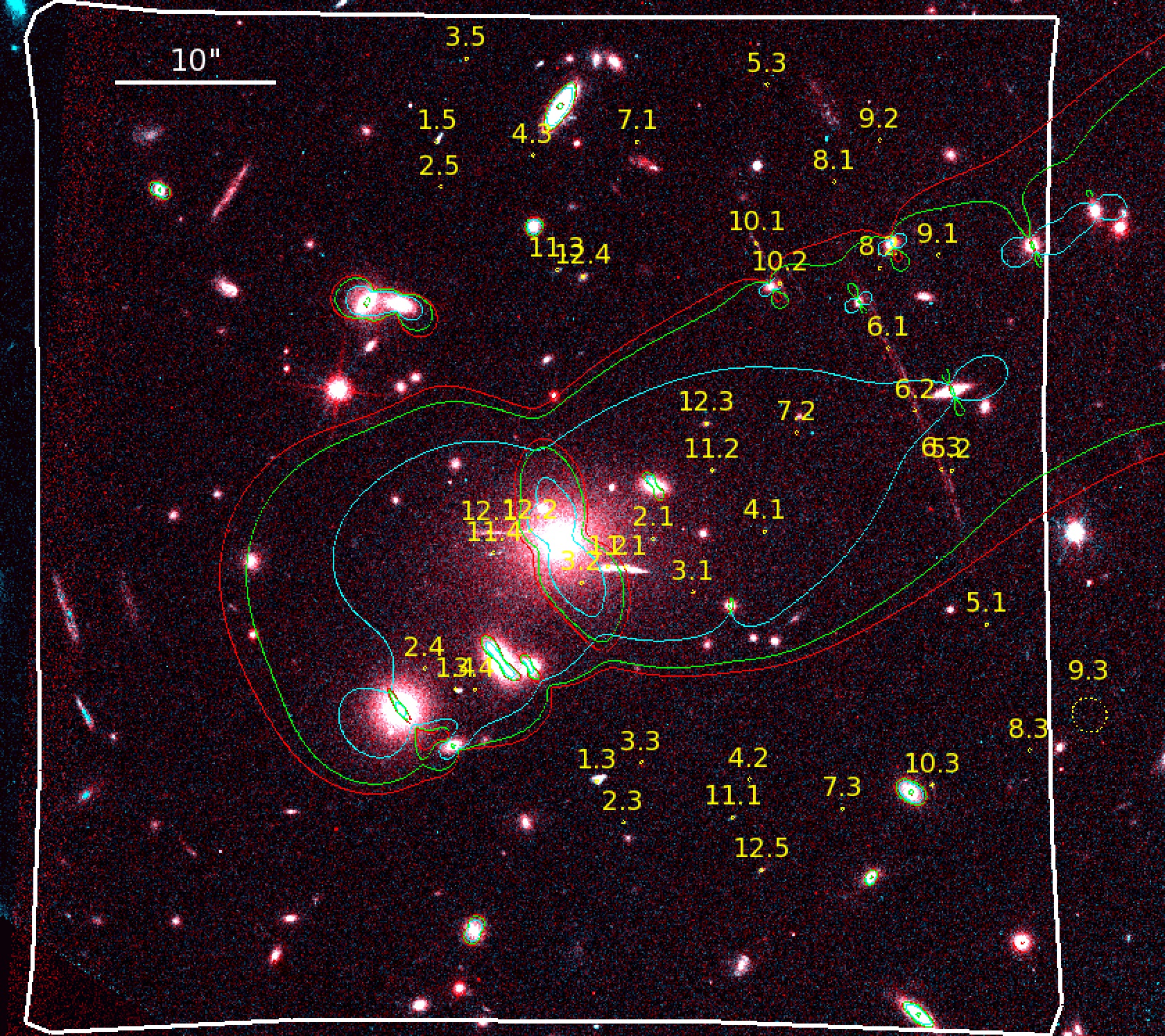}\\
\caption{Overview of all multiple-image systems (in yellow) identified over a Hubble color image combining the 
F606W and F814W bands (north up, east left).  The white polygon delineates the 
1x1 arcmin$^2$ region covered by the MUSE observations used in this paper. 
The cyan, green and red lines show the critical lines predicted by our revised 
mass model (Sect. \ref{modelling}) at $z=1.46$ (system 6), 
$z=3.5077$ (systems 1 and 2) and $z=6.4085$ (system 9).}
\label{hstimage}
\end{figure}

An early analysis of these images revealed a 5-imaged strongly lensed system identified 
by its morphology, and measured by \citet{Christensen2012} at $z=3.5077$ using XShooter (images 1.1 to 1.5  
in Figure \ref{hstimage}). Additional candidate radial arcs were discovered around the SE sub cluster, 
showing the potential for confirming more multiple images in this region.

For this reason, SMACSJ2031 was targeted as part of MUSE commissioning (ESO program 60.A-9100(B)) 
to demonstrate the power of MUSE once combined with lensing magnification. 
A total of 10.4 hours of observations were obtained in nominal (WFM-NOAO-N) mode between Apr 30th and May 7th 2014, and the 1x1 arcmin$^2$ 
field-of-view was centred on the SE cluster to cover the predicted locus of multiple images (Figure \ref{hstimage}). 
A small random dithering pattern (within a 0.6" box) and 90 degree rotations of the instrument were performed 
between each exposure of 1200 sec. 

The MUSE data were reduced and combined into a single datacube 
using the MUSE data reduction software (Weilbacher et al. in prep.), and further processed with the ZAP 
software (Soto et al. in prep.) which suppresses residuals of sky subtraction using a principal component analysis. 
Flux calibration was performed based on observations of the spectrophotometric standard HD49798 (O6 star) and GD108 (subdwarf B star). The final datacube is sampled at a pixel scale of 0.2" and covers the wavelength range 4750-9350 \AA\ with a dispersion of 1.25 \AA\ per pixel and a resolution increasing from R$\sim$2000 to R$\sim$4000 between the blue and red end. The seeing varied from 0.6 to 1.2'' between each exposure and is measured to be $\sim0.95$'' in the final combined datacube based on the brightest star in the field. More details about the data reduction and quality of the final products will be given in Patricio et al. (to be submitted).

\begin{figure*}
\begin{center}
\includegraphics[width=\textwidth]{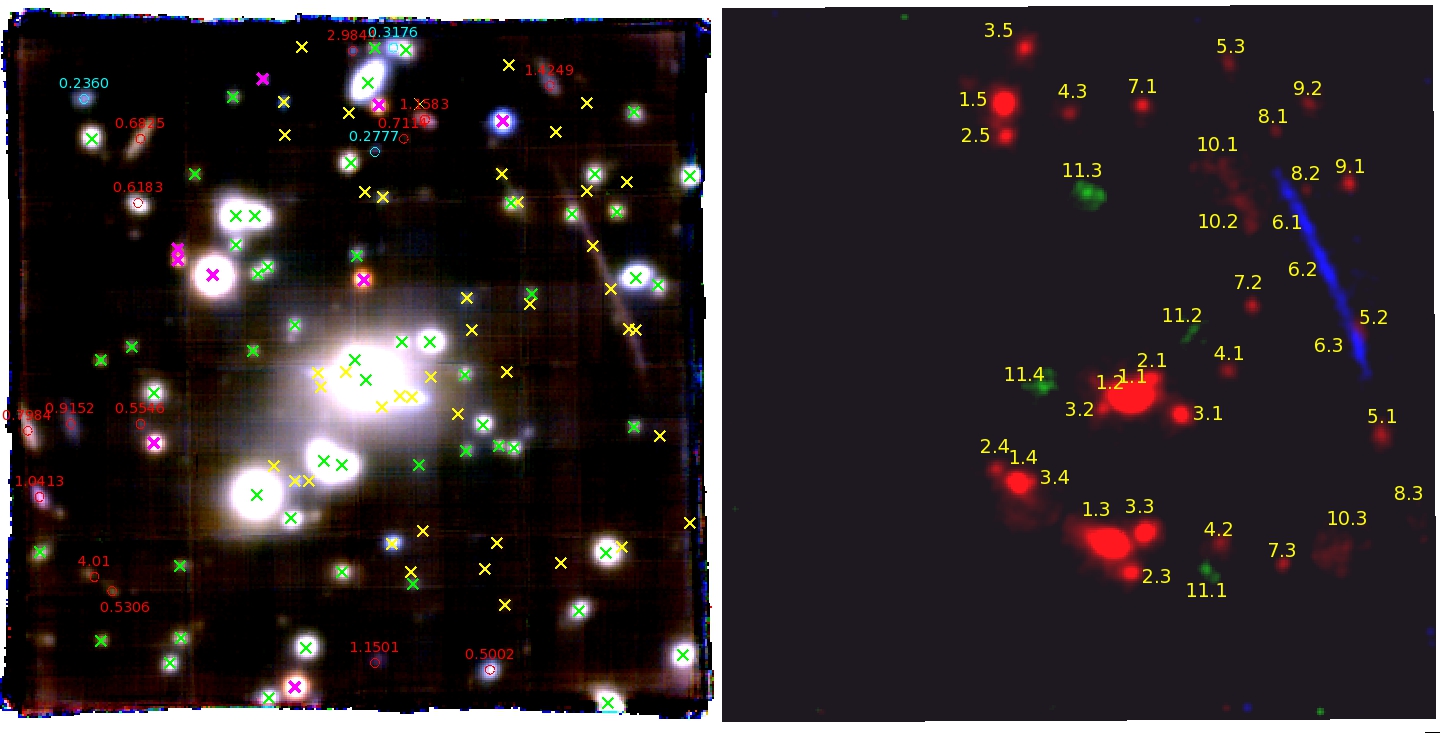}
\caption{Overview of all redshifts measured in MUSE observations. The left panel shows a MUSE  color image produced by collapsing 
three contiguous wavelength ranges in the whole cube. The green crosses identify cluster members confirmed by our spectroscopy. Cyan 
and red circles show the redshifts of foreground and single image background sources, respectively, while the yellow crosses mark 
the location of multiple images. The right panel shows a combination of all narrow band images showing the emission lines in multiply 
imaged systems 1 to 12. Lyman-$\alpha$ emitters are presented in red, while [OII] and CIII] emitters are shown in blue and green, respectively.}
\label{muse}
\end{center}
\end{figure*}

\section{Data Analysis}
\label{dataanalysis}

We searched the MUSE datacube for spectral features, both for continuum-detected sources and 
for objects dominated by nebular emission lines. We first produced ``white-light''  and color images 
 by collapsing the cube over its full wavelength range or in the blue/green/red wavelengths (Fig. \ref{muse}, left panel). 
 We ran SExtractor \citep{1996A&AS..117..393B} to identify objects in the white-light image 
 and measured their continuum levels at 3 wavelengths. The spectra of these continuum sources were extracted 
 using the spatial extent measured in the white light image.
 
In addition, we produced pseudo narrow-band images by performing a running average of 5 adjacent wavelength planes (6.25\AA\ width) in the 
 whole cube, using a weighting scheme to enhance the detection of typical $\sigma\sim100$ km/s emission lines. We 
 ran SExtractor again in each of these narrow-band images, and compared the detections with the previous continuum measurements
  to identify emission lines associated with continuum sources and isolated emission lines. Examples of continuum-subtracted narrow-band detections 
  are given in Fig. \ref{muse} (right panel). All emission lines are well-detected and resolved over multiple wavelength planes throughout the 
  cube.

\begin{figure*}
\includegraphics[width=\textwidth]{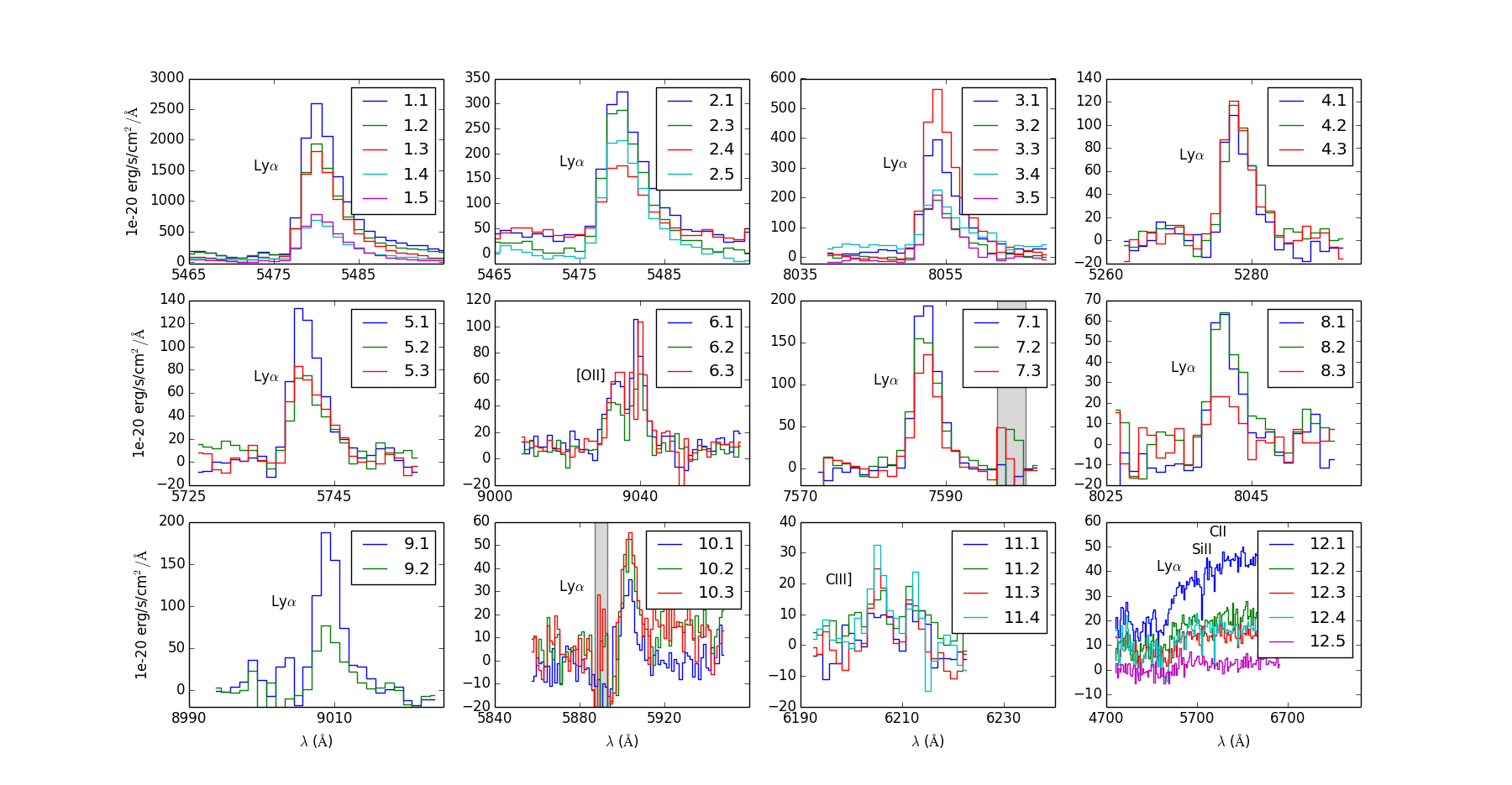}
\caption{Extracted spectra of the strongest emission line in all identified multiple images, grouped by lensing system (in reading order). One 
can clearly see the asymmetric profile of all Lyman-$\alpha$ emission lines, as well as the doublet for [OII] and CIII]. In system 12 the 
main feature is the Lyman-$\alpha$ break as well as rest-frame UV absorption lines (SiII 1304 \AA\ , CII 1334 \AA). The grey areas mark the location of strong sky line residuals.}
\label{specmosaic}
\end{figure*}

Redshifts were measured for both continuum and emission-line dominated sources by identifying a combination of emission and 
absorption line features. The large majority of continuum sources identified are formed by cluster members (53) and foreground sources (3), 
easily recognisable with K 3933 \AA\ and H 3969 \AA\ absorption features (green and cyan crosses respectively in the left panel of Fig. \ref{muse}). Background 
sources are all identified through their emission lines or Lyman-$\alpha$ break, in particular [OIII] 5007 \AA\ at $z<0.8$, [OII] 3727 \AA\ at $<1.5$, but for the large majority with  Lyman-$\alpha$ 1216 \AA\ at 
$z>2.8$ (Table \ref{mutable}). As Lyman-$\alpha$ is usually the only spectral feature visible in the MUSE wavelength range, we rely on its 
typical asymmetric shape and the absence of a doublet or other emission lines to ascertain the redshift (Fig. \ref{specmosaic}). We also measured the redshifts of 
candidate arcs and multiple images visible in HST images, through [OII] at $z=1.46$ (system 6), CIII] 1907,1909 \AA\ at $z=2.2556$ (system 11) or a Lyman-$\alpha$ break at $z=3.41$ (system 12).

We searched through the redshifts measured in the cube for multiple images based on a preliminary mass modelling of the 
cluster core, which used only the 5-imaged system at $z=3.5077$ \citep{Christensen2012}. We found reliable connections 
for 43 sources forming a total of 12 systems (combining continuum and emission-line dominated sources), with the predictions from the model 
always lying within 1 arcsec from the detection. We then compared the extracted spectra for all 12 multiple images and found 
a very good match, as seen in Fig. \ref{specmosaic}. Together with our improved mass model (see Sect. \ref{modelling}) this 
clearly confirms these sources as multiple systems. 

\begin{table}
\input{multableflux.tex}
\input{bkgtableflux.tex}
 \vspace{-0.3cm}
\caption{
\label{mutable}Multiple image systems (top part) and other single image lensed sources (bottom part). From left 
to right we give their positions (J2000.0), measured redshifts, magnifications, main line feature detected and corresponding line flux.
 Ly$\alpha$b refers to the 
Lyman-$\alpha$ absorption break. Positions marked with brackets are not detected with MUSE but are predicted by the 
lens model (see text for details). }
\end{table}

\section{Cluster mass modelling}
\label{modelling}

Following our previous work on cluster mass modelling \citep{2007ApJ...668..643L,Jauzac2014a}, we used the locations 
and redshifts of the 12 multiple systems identified to constrain a parametric mass model of the cluster using the Lenstool 
\citep{2007NJPh....9..447J} software\footnote{publicly available \href{http://projects.lam.fr/projects/lenstool/wiki}{on the dedicated web page}}.
In short, we adopt a dual pseudo-isothermal elliptical mass distribution (dPIE, also known as truncated PIEMD, see  \citealt{Ardis}),  which corresponds to an isothermal profile with two characteristic radii: a core radius r$_{\rm core}$ (producing a flattening of the mass distribution at the centre), and a cut radius r$_{\rm cut}$ (producing a drop-off of the mass distribution on large scales). Our preliminary mass model of SMACSJ2031, which we used 
in \citet{Christensen2012} to derive magnification factors, assumed a single cluster-scale dPIE component centred on the SE cluster as well as individual 
galaxy-scale dPIE profiles on each cluster member following a scaling relation with luminosity (see e.g. \citealt{2010MNRAS.404..325R}).

With the addition of new multiple images, we refine this mass model by incorporating a second cluster-scale mass component centred on the NW 
cluster, located 1.2 arcmin away. This extends the critical line in the NW-SE direction (Fig. \ref{hstimage}) and gives a much better match 
with the new systems 6 to 10 located in this region. One can clearly see the effect of this extension of the critical line towards the upper-right corner of the MUSE 
field-of-view, with an absence of multiple images identified in the lower-left corner. The best fit parameters of the new mass model are presented in Table 
\ref{modelparams}.

\begin{table*}
\begin{tabular}{llllllll}
\hline
Potential & $\Delta\alpha$ & $\Delta\delta$ & $e$ & $\theta$ & r$_{\rm core}$ & r$_{\rm cut}$ & $\sigma$ \\
   & [arcsec] & [arcsec] & & [deg] & [kpc] & [kpc] & [km/s] \\
\hline
\input{modelresults.tex}
\end{tabular}
\caption{\label{modelparams} Best-fit model parameters of the cluster-scale dPIE profiles (SE and NW subclusters) and the scaling 
relation of cluster members (shown for a L$^*$ galaxy. From left 
to right: center in arcsecs from the origin fixed at the location of the brightest cluster galaxy ($\alpha$=20:35:11.835,$\delta$=-40:27:08.47), ellipticity and position angle, core and cut radii, central 
velocity dispersion. Values in brackets are fixed in the modelling.}
\end{table*}

\begin{figure}
\includegraphics[width=0.45\textwidth]{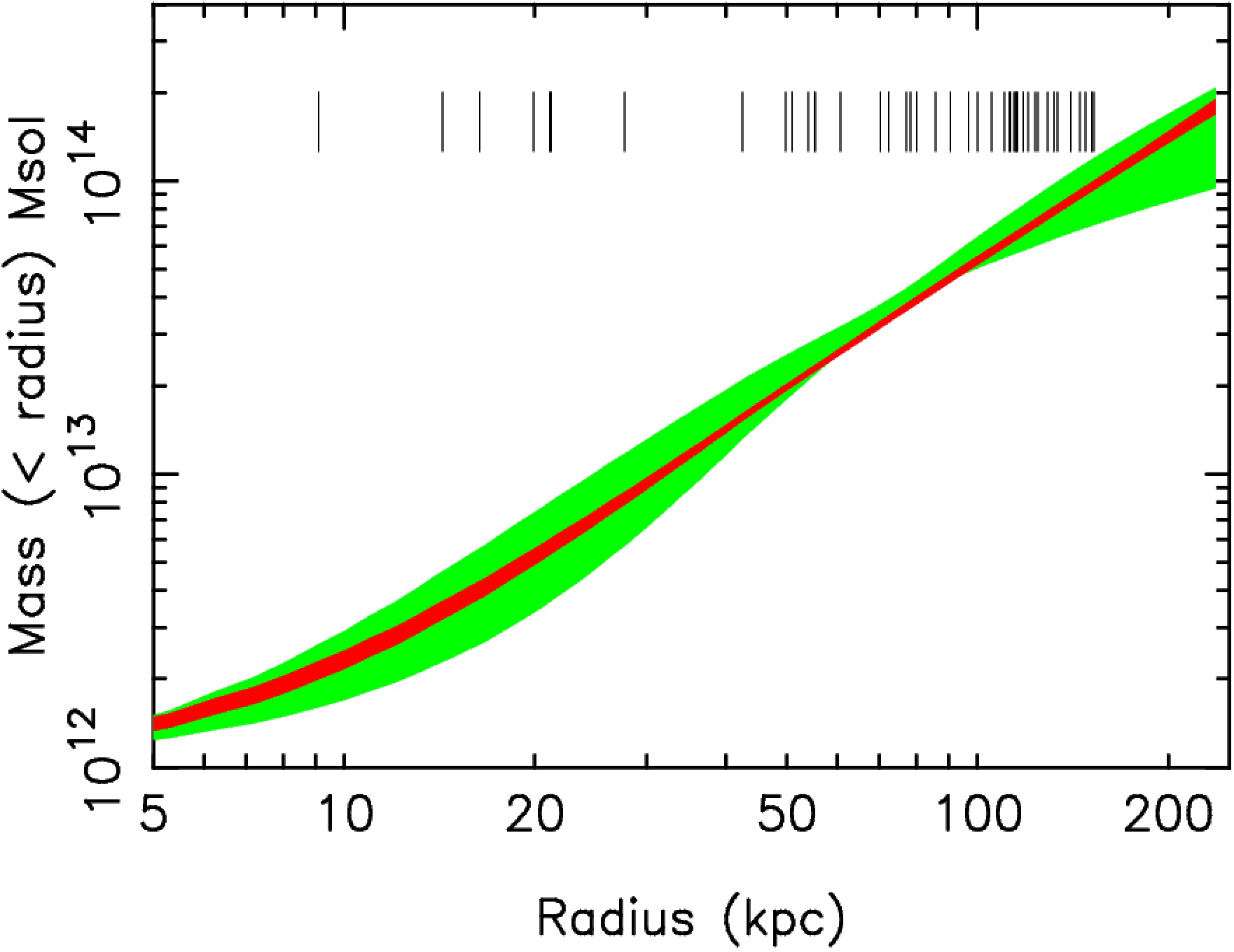}
\caption{Projected cluster mass profile integrated over the radial distances from the center of the SE cluster (red curve). 
The radial distance of all multiple images used as strong-lensing constraints are marked with black ticks. 
We overplot in green the range of mass profiles allowed by the previous mass model (which only used constraints from 
system 1), showing the gain in precision obtained with the MUSE data.
}
\label{profile}
\end{figure}

The two-component mass model now reproduces all 12 multiple systems with an rms of 0.41" in the image plane. Although the 
centroid precision of the emission lines is much better than this for the MUSE data (typically 0.1") such a scatter is expected because of 
additional lensing structures along the line of sight (see the discussion in \citealt{2010Sci...329..924J}). An illustration of the ability to reproduce 
the systems is shown in the critical lines presented in Fig. \ref{hstimage} as we expect a symmetry of multiple images against the 
corresponding critical line. The projected cluster mass profile (radially integrated from the location of the 
brightest galaxy in the SE cluster) is presented in Fig. \ref{profile}, together with the strong lensing constraints. We can see that the 
profile is best constrained within 10-150 kpc by the large number of multiple images. The precision obtained on the integrated 
mass in this region is a few percent, about 5 times better than the previous model.

We use this mass model to estimate the magnifications for all background sources, and report the values in Table \ref{mutable}. Errors on the magnification factors are estimated for each source by producing 500 mass models sampling the posterior probability distribution of the dPIE parameters, and measuring the distribution of  magnifications at this location. Thanks to the high density of constraints in the cluster core, the relative error on magnification factors are typically $5-30\%$, which is similar to the precision reached in the Hubble Frontier Fields 
\citep{Jauzac2014a}.

We also check that all of the 14 other background sources identified (s1 to s14 in Table \ref{mutable}, ranging from $z=0.5$ to $z=4.01$) 
are indeed predicted to be single images by this model, as they are all located in the SE corner of the MUSE field-of-view. Three 
counter images identified are not detected in the observations for the following reasons: image 9.3 is predicted to lie outside of the 
MUSE field of view, and images 2.2 and 11.5 are predicted to be located under the brightest cluster member, thus suffering from strong 
cluster light contamination.

To summarise, the sensitivity, resolution and wide field-of-view of MUSE, when coupled with the strong 
lensing region of a cluster core, form the perfect combination to detect numerous lensing systems and use 
them as constraints to measure the cluster mass profile. Further, more detailed analysis could include the 
resolved constraints from cluster members in the cluster core, for which we can measure the individual 
velocity dispersions with the MUSE data.

In addition, accounting for the magnification factor, we show here that we can reliably measure the redshifts 
for Lyman-$\alpha$ emitters that are intrinsically as faint as $2\times10^{-19}$ erg/s/cm$^{2}$ (images 8.1 and 8.2) and up to $z=6.4$ (system 9). The 
multiplicity of these sources in the strong lensing regime helps to confirm these faint lines to be Lyman-$\alpha$. 
We expect that targeting lensing cluster with MUSE will result in the detection of a large number of such 
faint emitters at high redshift.

\section*{Acknowledgments}
We thank Eric Emsellem for his comments on the manuscript. JR and VP acknowledges support from the ERC starting grant CALENDS. JR acknowledges support from the CIG grant 294074. 
Based on observations made with ESO Telescopes at the La Silla Paranal Observatory under programme ID 60.A-9100(B). We thank all the staff at 
Paranal Observatory for their valuable support during the commissioning of MUSE.  Also based on observations made with the NASA/ESA Hubble 
Space Telescope (PID 12166 and 12884). 


\bibliographystyle{mn2e}
\bibliography{references}




\label{lastpage}

\end{document}

%% file: multableflux.tex
\begin{tabular}{llllrll}
ID & $\alpha$ (hrs) & $\delta$ (deg) & $z_{\rm spec}$ & $\mu$ & line & flux \\
   & 20:31: & -40:37: & & & &\hspace{-0.5cm}10$^{-18}$ erg/s/cm$^2$ \\
\hline 
 1.1 & 52.90 & 32.5 & 3.508 & $19.9\pm1.9$ & Ly$\alpha$ & $185\pm 5$\\
 1.2 & 52.99 & 32.5 & 3.508 & $11.7\pm1.2$ & Ly$\alpha$& $169\pm 4$\\
 1.3 & 53.05 & 45.8 & 3.508 & $6.8\pm0.2$ & Ly$\alpha$& $270\pm 3$\\
 1.4 & 53.83 & 40.1 & 3.508 & $2.5\pm0.1$ & Ly$\alpha$& $42\pm 0.6$\\
 1.5 & 53.93 & 06.0 & 3.508 & $3.5\pm0.1$ & Ly$\alpha$& $110\pm 2$\\
 2.1 & 52.74 & 30.7 & 3.508 & $4.7\pm0.2$ & Ly$\alpha$& $22.4\pm 0.6$\\
 2.2 & [53.12] & [31.1] &    &    & & \\
 2.3 & 52.91 & 48.4 & 3.508 & $4.7\pm0.1$ & Ly$\alpha$& $21.1\pm 0.8$\\
 2.4 & 53.99 & 38.8 & 3.508 & $2.7\pm0.1$ & Ly$\alpha$& $10.4\pm 0.6$\\
 2.5 & 53.90 & 08.9 & 3.508 & $3.9\pm0.1$ & Ly$\alpha$& $16.4\pm 0.7$\\
 3.1 & 52.53 & 34.1 & 5.623 & $5.3\pm0.2$ & Ly$\alpha$& $37.6\pm 1.4$\\
 3.2 & 53.14 & 33.5 & 5.623 & $2.6\pm0.1$ & Ly$\alpha$& $17.7\pm 1.1$\\
 3.3 & 52.81 & 44.6 & 5.623 & $8.2\pm0.3$ & Ly$\alpha$& $56.6\pm 1.8$\\
 3.4 & 53.72 & 40.1 & 5.623 & $2.2\pm0.1$ & Ly$\alpha$& $11.6\pm 0.9$\\
 3.5 & 53.76 & 00.9 & 5.623 & $3.3\pm0.1$ & Ly$\alpha$& $24.8\pm 1.4$\\
 4.1 & 52.14 & 30.3 & 3.340 & $4.1\pm0.1$ & Ly$\alpha$& $8.3\pm 0.7$\\
 4.2 & 52.22 & 45.7 & 3.340 & $5.0\pm0.1$ & Ly$\alpha$& $12.6\pm 1.0$\\
 4.3 & 53.40 & 06.9 & 3.340 & $5.2\pm0.1$ & Ly$\alpha$& $9.9\pm 0.8$\\
 5.1 & 50.92 & 36.1 & 3.723 & $9.1\pm0.2$ & Ly$\alpha$& $14.3\pm 1.0$\\
 5.2 & 51.11 & 26.5 & 3.723 & $18.1\pm0.8$ & Ly$\alpha$& $6.8\pm 0.7$\\
 5.3 & 52.12 & 02.5 & 3.723 & $5.1\pm0.1$ & Ly$\alpha$& $6.5\pm 0.7$\\
 6.1 & 51.46 & 18.9 & 1.425 & $19.5\pm1.0$ & [OII]& $582\pm 13$\\
 6.2 & 51.31 & 22.8 & 1.425 & $11.8\pm0.5$ & [OII]& $578\pm 12$\\
 6.3 & 51.17 & 26.4 & 1.425 & $18.2\pm0.6$ & [OII]& $553\pm 11$\\
 7.1 & 52.83 & 06.1 & 5.240 & $4.9\pm0.1$ & Ly$\alpha$& $14.5\pm 0.8$\\
 7.2 & 51.96 & 24.2 & 5.240 & $4.2\pm0.1$ & Ly$\alpha$& $8.1\pm 0.6$\\
 7.3 & 51.71 & 47.6 & 5.240 & $4.1\pm0.1$ & Ly$\alpha$& $6.7\pm 0.5$\\
 8.1 & 51.75 & 08.6 & 5.613 & $15.0\pm0.5$ & Ly$\alpha$& $3.4\pm 0.4$\\
 8.2 & 51.51 & 13.9 & 5.613 & $14.0\pm0.6$ & Ly$\alpha$& $2.7\pm 0.3$\\
 8.3 & 50.69 & 43.9 & 5.613 & $4.1\pm0.1$ & Ly$\alpha$& $0.4\pm 0.1$\\
 9.1 & 51.19 & 13.1 & 6.408 & $14.8\pm0.9$ & Ly$\alpha$& $10.3\pm 0.8$\\
 9.2 & 51.51 & 06.0 & 6.408 & $13.8\pm0.6$ & Ly$\alpha$& $5.5\pm 0.7$\\
 9.3 & [50.36] & [41.7] &    &       & & \\
 10.1 & 52.18 & 12.4 & 3.856 & $14.1\pm0.4$ & Ly$\alpha$& $4.0\pm 0.8$\\
 10.2 & 52.05 & 14.9 & 3.856 & $4.1\pm0.1$ & Ly$\alpha$& $9.6\pm 1.3$\\
 10.3 & 51.22 & 46.1 & 3.856 & $6.1\pm0.2$ & Ly$\alpha$& $8.9\pm 1.3$\\
 11.1 & 52.31 & 48.1 & 2.256 & $3.7\pm0.1$ & CIII]& $2.5\pm 0.8$\\
 11.2 & 52.42 & 26.5 & 2.256 & $3.4\pm0.1$ & CIII]& $1.4\pm 0.6$\\
 11.3 & 53.27 & 14.0 & 2.256 & $5.6\pm0.1$ & CIII]& $9.6\pm 2.6$\\
 11.4 & 53.62 & 31.6 & 2.256 & $4.9\pm0.3$ & CIII]& $6.3\pm 1.4$\\
 11.5 & [53.27] & [30.8] &    &       & & \\
 12.1 & 53.64 & 30.3 & 3.414 & $5.6\pm0.4$ & Ly$\alpha$b& \\
 12.2 & 53.42 & 30.2 & 3.414 & $4.3\pm0.3$ & Ly$\alpha$b& \\
 12.3 & 52.45 & 23.6 & 3.414 & $4.4\pm0.1$ & Ly$\alpha$b& \\
 12.4 & 53.13 & 14.4 & 3.414 & $7.3\pm0.2$ & Ly$\alpha$b& \\
 12.5 & 52.15 & 51.4 & 3.414 & $3.2\pm0.1$ & Ly$\alpha$b& \\
\hline 

%% file: bkgtableflux.tex
s1  & 55.37 & 49.2 & 4.010  & $3.5\pm0.1$ & Ly$\alpha$ & $12.8\pm 1.1$\\
s2  & 53.37 & 01.2 & 2.984 & $3.4\pm0.1$ & Ly$\alpha$ & $38.7\pm 1.5$\\
s3  & 51.80 & 04.1 & 1.425 & $4.2\pm0.2$ & [OII]  & $75.6\pm 2.3$ \\
s4  & 52.79 & 07.4 & 1.158 & $3.1\pm0.1$ & [OII]  & $60.2\pm 2.0$ \\
s5  & 53.16 & 56.3 & 1.150 & $2.1\pm0.1$ & [OII]  & $19.1\pm 1.7$ \\
s6  & 55.85 & 41.7 & 1.041 & $2.3\pm0.1$ & [OII]  & $48.7\pm 1.7$\\
s7  & 55.61 & 35.0 & 0.9152 & $2.5\pm0.1$ & [OII]  & $62.7\pm 1.4$\\
s8  & 55.94 & 35.6 & 0.7984 & $2.0\pm0.1$ & [OII] & $100.7\pm1.3$ \\
s9  & 52.97 & 09.0 & 0.7114 & $2.2\pm0.1$ & [OII] & $3.9\pm 0.8$\\
s10 & 55.06 & 09.2 & 0.6825 & $1.7\pm0.1$ & [OII] & $75.7\pm 1.9$ \\
s11 & 55.07 & 15.2 & 0.6183 & $1.8\pm0.1$ & [OII] & $47.8\pm 1.9$ \\
s12 & 55.05 & 35.0 & 0.5546 & $1.4\pm0.1$ & [OII] & $4.2\pm 1.0$ \\
s13 & 55.27 & 50.1 & 0.5306 & $1.9\pm0.1$ & K,H   & \\
s14 & 52.28 & 57.2 & 0.5002 & $1.4\pm0.1$ & [OII] & $112.8\pm3.9$ \\
\hline
\end{tabular}

%% file: modelresults.tex
SE  & $  0.5^{+  0.2}_{ -0.1}$ & $ -0.5^{+  0.2}_{ -0.2}$ & $ 0.29^{+ 0.04}_{-0.04}$ & $  3.0^{+  2.7}_{ -3.7}$ & $29^{+2}_{-2}$ & $[1000]$ & $614^{+24}_{-22}$ \\
NW  & $ 62.2^{+  0.5}_{ -0.5}$ & $ 25.3^{+  1.0}_{ -0.9}$ & $ 0.47^{+ 0.08}_{-0.06}$ & $  6.2^{+  6.4}_{ -1.6}$ & $124^{+15}_{-13}$ & $[1000]$ & $1062^{+26}_{-54}$ \\
L$^{*}$ galaxy &  & & & & $[0.15]$ & $32^{+6}_{-5}$ & $117^{+7}_{-8}$\\
\hline